# The Untapped Potential of Smart Charging: How EV Owners Can Save Money and Reduce Emissions Without Behavioral Change

*Yash Gupta*[*1], William Vreeland[1], Andrew Peterman[1], Coley Girouard[1], Brian Wang[1]*
*[1]Rivian Automotive, Palo Alto, CA, USA*
**[*]Email: yashgupta@rivian.com; yashg2607@gmail.com**

## Abstract

The transportation sector is the single largest contributor to US emissions and the second largest globally. Electric vehicles (EVs) are expected to represent half of global car sales by 2035, emerging as a pivotal solution to reduce emissions and enhance grid flexibility. The electrification of buildings, manufacturing, and transportation is expected to grow electricity demand substantially over the next decade. Without effectively managed EV charging, EVs could strain energy grid infrastructure and increase electricity costs. Drawing on de-identified 2023 EV telematics data from Rivian Automotive, this study found that 72% of home charging commenced after the customer plugged in their vehicle regardless of utility time of use (TOU) tariffs or managed charging programs. In fewer than 26% of charging sessions in the sample, EV owners actively scheduled charging times to align or participate in utility tariffs or programs. With a majority of drivers concurrently plugged in during optimal charging periods yet not actively charging, the study identified an opportunity to reduce individual EV owner costs and carbon emissions through smarter charging habits without significant behavioral modifications or sacrifice in user preferences. By optimizing home charging schedules within existing plug-in and plug-out windows, the study suggests that EV owners can save an average of $140 annually and reduce the associated carbon emissions of charging their EV by as much as 28%.

## Introduction

Electrifying transportation plays a crucial role in combatting climate change and reducing global reliance on fossil fuels [1,2]. The United States Environmental Protection Agency estimates that the transportation sector accounts for 28% of $CO_2$ emissions in the US [3] and 16.2% globally [4]. The International Energy Agency (IEA) reports that nearly one in five cars sold in 2023 were electric and projects that half of all global car sales will be electric by 2035 based on current climate policies [5]. Transitioning away from Internal Combustion Engine (ICE) vehicles has the potential to avoid over 2 gigatons of greenhouse gas emissions and reduce oil demand by more than 10 million barrels per day by 2035 [5].

Widespread EV adoption presents both opportunities and challenges for the US energy grid. EV electricity demand has the potential to reach up to 14% of total electricity demand in the US by 2035, up from 0.6% today [5]. While EVs can reduce electricity costs, support renewable energy integration, and enhance grid flexibility [6-13], uncoordinated charging can increase peak loads which strains grid infrastructure and leads to higher electricity costs [14-17]. The potential for unmitigated EV charging was found to be especially concerning under scenarios in which EV adoption occurs rapidly [18]. Grid planners forecast a 38 GW increase in peak electricity demand within five years, driven by cross sectoral electrification and growth in energy-intensive AI services – equivalent to adding another California to an already overburdened grid [19]. To date, EVs have been a valuable grid resource in many parts of the country. Between 2011 and 2021, EV drivers contributed over $3 billion more in revenues than their associated grid costs [13]. This surplus helps stabilize electricity rates for everyone and enables utilities to respond to growing electricity demands and cost pressures.

According to the IEA, most EV charging currently occurs at home – 83% in the US [5]. Residential charging presents unique opportunities to support the energy grid compared to public charging. Due to long dwell times at home, sometimes spanning multiple days, in which charging can occur and still meet user needs, our analysis revealed that, on average, vehicles at home remain plugged in twice as long as needed to achieve the owner's desired state of charge. EV owners typically initiate charging immediately upon returning home, leading to suboptimal charging patterns that result in unnecessarily higher energy costs, grid inefficiency and increased emissions. The data revealed that 72% of home charging begins immediately after the customer plugged in their vehicle upon arrival, regardless of time-of-use (TOU) rates or incentives designed to shift charging to more optimal times for the grid. Approximately 31% of sessions began during peak hours (4 pm to 9 pm) and in fewer than 26% of sessions, EV owners actively scheduled charging times to align with utility tariffs or programs. This finding is consistent with prior research in which 70% of users charged immediately upon arrival and 30% delayed charging from 15 minutes to 5 hours post-arrival [20].

Utilities manage energy load from EVs through two primary approaches:

1. Passive TOU Rate Designs: A passive pricing mechanism that incentivizes beneficial charging times – familiarly known as peak and off-peak pricing.
2. Active Grid Management Programs: In addition to TOU rates, programs enable dynamic utility control over charging processes to address both system-wide and localized grid congestion, offering compensation to participants for their flexibility.

Both mechanisms aim to align charging patterns with optimal grid operations by influencing consumer behavior. TOU rates offer a largely static approach, while managed charging implements a more active and targeted strategy.

However, EV owners' behavior often deviates from what is best for the grid. Most customers fail to align charging with off-peak hours without incentives [21] and only 51% of EV owners are even aware of residential EV utility incentive programs with a minority (35%) who regularly schedule their charging times [22]. It is further compounded by the delayed feedback on electricity bills, which typically arrive at least a month after the charging occurs. This imposes significant cognitive demands on consumers, leading them to make sub-optimal charging decisions based on heuristics rather than complex rate comparisons and charging schedule adjustments. Additionally, many consumers prioritize convenience and flexibility over cost optimization, preferring to maintain higher battery levels for unexpected travel needs [23]. These findings highlight a significant gap between optimal and actual charging behaviors.

EV makers can bridge this divide by leveraging utility rates and other grid signals to provide real-time information and feedback through smart charging customer interfaces. As shown in Table 1, various studies define and quantify the impact and potential benefits of smart charging.

***Table 1:*** *A non-exhaustive list of studies found in the literature focused on smart charging and its benefits*

| Study | Smart Charging Optimizations | Results & Conclusions |
|---|---|---|
| [24] | Aligning EV charging times with grid benefits, off-peak hours, and renewable integration. | Reduced charging costs by 30%, grid operational costs by 10%, and renewable curtailment by 40%. |
| [25] | Two smart charging strategies: loss-optimal (charging when there is less demand) and cost-optimal (charging when electricity costs are lower) by modeling a real-world distribution network over a year using hourly resolution data from Stockholm, assuming home-office-home driving patterns with 20-80% SOC limits and fully controllable chargers with variable power output. | Enabled grid loss reductions of up to 35% and 61% reductions in charging costs compared to uncontrolled charging. |
| [26] | Smart charging through solar PV integration, optimizing costs by using excess solar power generated to charge EVs under two scenarios: smart16 in which the vehicle isn't at home during working hours from 8 AM to 4 PM and smart24 with 24-hour availability. | Households achieved over 50% cost savings for EV charging when charging times and durations were optimized. |
| [27] | Optimize charging windows based on multiple inputs: vehicle state of charge, projected unplug time, utility pricing, grid conditions, renewable energy availability, and user preferences (cost vs. renewable energy). | 32% reductions in carbon emissions and $325 in estimated grid value per vehicle annually. |
| [28] | Emissions optimization approach that incorporated additional flexibility periods of 6 to 24 hours beyond the original charging session completion time. | Reduced carbon emissions by 8% to 14%, with potential reductions up to 43% in states such as California with higher renewable energy penetration |
| [29] | Utilizing emissions signals to synchronize charging with cleaner energy periods (but ignoring the TOU rate structure, thus not accounting for costs) while ensuring full battery charge completion with two charging windows: workplace (9 AM to 5 PM) and home (7 PM to 7 AM). | 18% reductions in annual greenhouse gas emissions |

Prior studies demonstrate substantial benefits of smart charging, such as lower costs for grid operators and EV owners, and reductions in system-wide carbon emissions. However, implementing these strategies often assumes unrealistic vehicle availability and charging flexibility. Many studies assume extended home availability without taking into account the EV owner's daily routine, which is a key limitation to smart charging adoption. Research on German EV adopters confirms that while grid stability and renewable integration are key motivators, the desire for flexible mobility often outweighs cost and emissions benefits [30]. This study quantifies the potential carbon and cost reductions of smart charging *without* behavioral modifications, demonstrating that significant benefits are possible even without altering customer habits. While behavioral modifications can yield greater benefits, this research highlights the readily achievable gains of smart charging without behavioral changes.

This study quantifies the potential benefits of smart charging by first minimizing electricity costs by charging during lower-priced electricity periods (off-peak hours) with carbon emission reduction as a secondary objective. This mode is referred to as smart charging hereafter and more details on the optimization framework described in the ***Methods*** section. For clarity, throughout this study, the term 'emissions' specifically refers to Scope 2 emissions (CO2e) from electricity consumption related to EV charging. Historical residential charging and vehicle telematics data are used to quantify two research questions:

1. To what extent and by how much can EV owners reduce their actual costs and associated carbon emissions from charging at home?
2. To what extent and by how much are these benefits achievable without behavioral modifications to EV owners' existing plug-in and plug-out schedules?

The study uses real-world, de-identified home charging data from Rivian Automotive – a US-based automaker building all-electric pick-up trucks, sport utility vehicles, and commercial vans.

Rivian charging telematics were combined with residential utility tariff information from Arcadia and time-varied grid marginal carbon emissions data from WattTime.

Rivian's telematics data provides detailed information on home charging sessions for customers who were opted-into data sharing for analytics purposes, including time of charge, state of charge, plug-in and plug-out times at user-designated "Home" locations. These data enable the historical assessment of potential savings under two scenarios: an unconstrained scenario, where charging is 'theoretically' optimized without time restrictions, and a constrained scenario, where optimization occurs within the actual plug-in times of each session. Figures 1 and 2 illustrate this, showing the distribution of plug-in session durations and the percentage of plug-in time spent charging, categorized by the time of day when the vehicle is plugged in. The unconstrained optimization calculates the historical maximum possible savings, while the constrained scenario quantifies what is achievable without behavioral changes (or when vehicles were found to be already plugged in and capable of charging).

To assess cost and emissions benefits, the study incorporated residential tariffs from various utilities using Signal API enabled by Arcadia [31] and marginal carbon emissions data for grid and sub-grid regions through the non-profit data provider, WattTime. Both datasets are crucial in enabling and evaluating the potential cost and emission savings from residential charging. A detailed discussion for each of these datasets can be found in the ***Supplementary Information*** section.

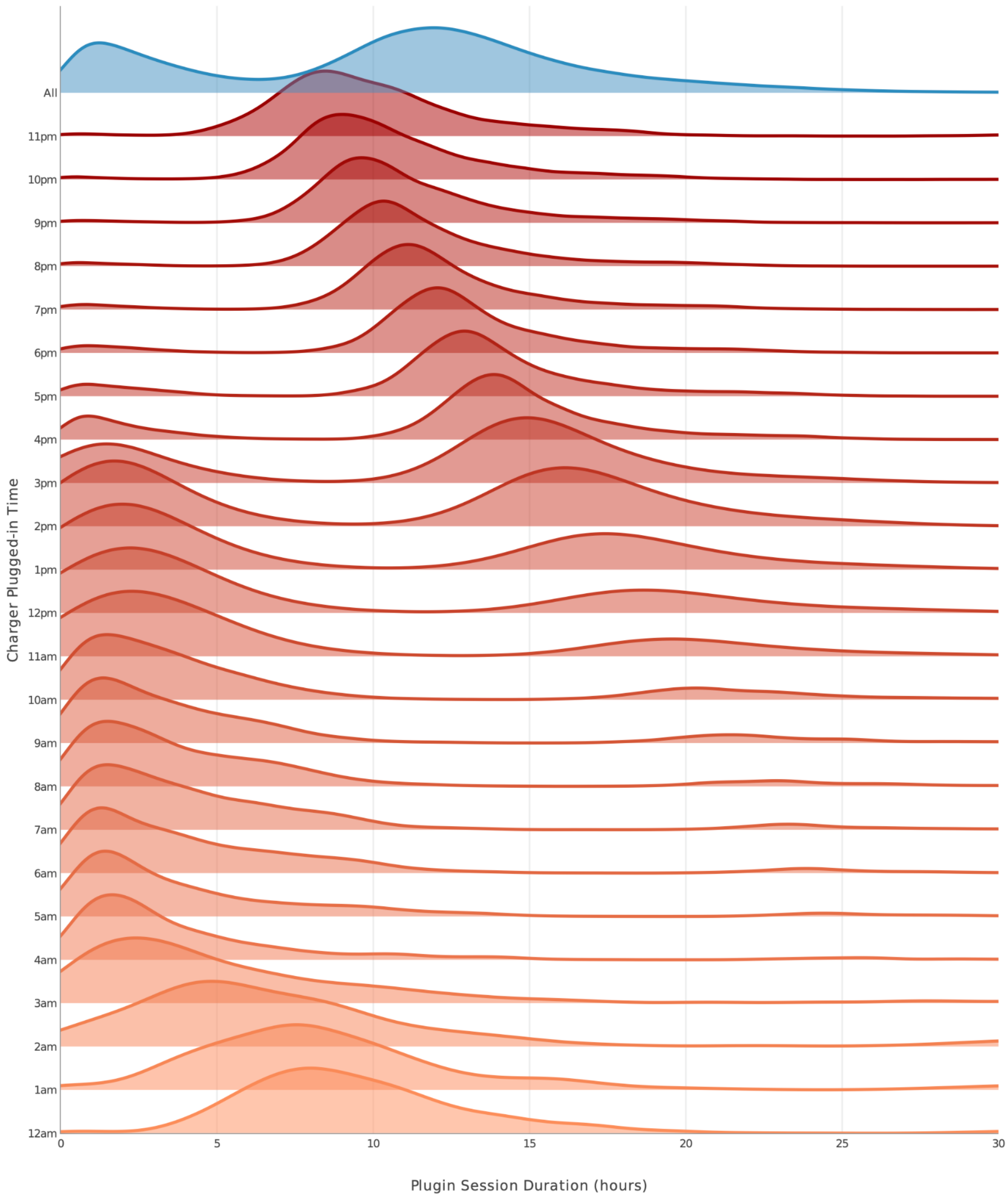

***Figure 1:*** *Distribution of plug-in session's duration for different charger plugin times during the day (mean = 11.2 hours; median = 11.4 hours)*

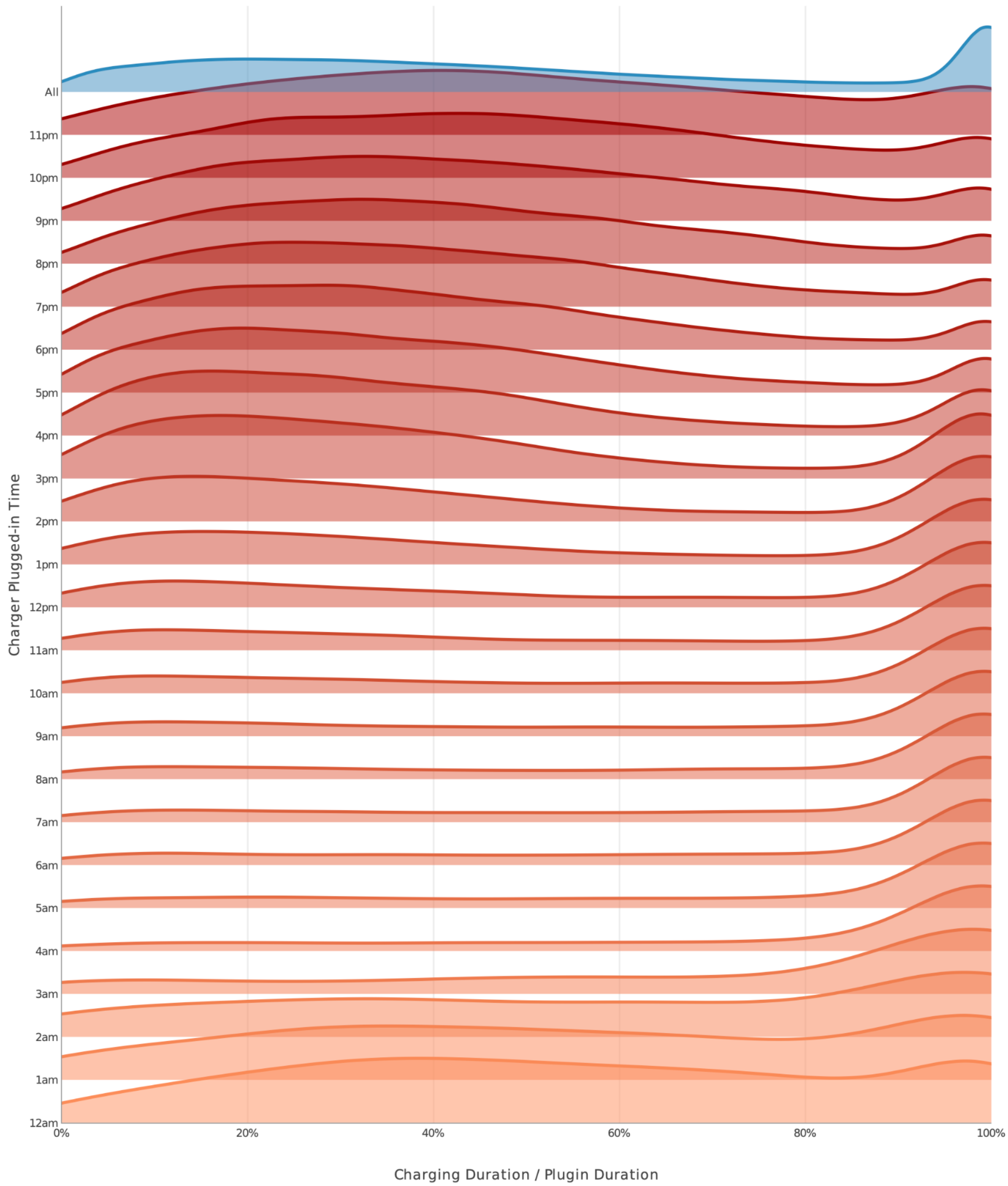

***Figure 2:*** *Distribution of percent of total plug-in duration spent charging for different charger plugin times during the day (mean = 49.7%; median = 43.1%)*

# Methods

**Optimization Framework:**

This study uses two retrospective optimization frameworks to assess the potential cost and carbon emission savings of residential EV charging:

1. Constrained Optimization Scenario: Charging is optimized only within the customer's actual plug-in times of each charging session. This approach optimizes all the charging sessions, single or multiple, within a single plug-in window and quantifies the potential benefits without any behavioral changes in customer charging.
2. Unconstrained Optimization Scenario: Charging is optimized without any time restrictions. This approach optimizes all the charging sessions on a given day (which can sometimes include multiple sessions), establishing an upper bound on potential savings.

Figure 3 illustrates these scenarios using data from two Rivian vehicles. The first one served by Pacific Gas & Electric Co. (PG&E) in the CAISO North grid region, and the second one served by Southern California Edison Co. (SCE) in the CAISO San Bernardino grid regions. The visualization demonstrates the optimization when restricted within the actual plug-in times of each charging session and cases where multiple sessions occur within a single plug-in window.

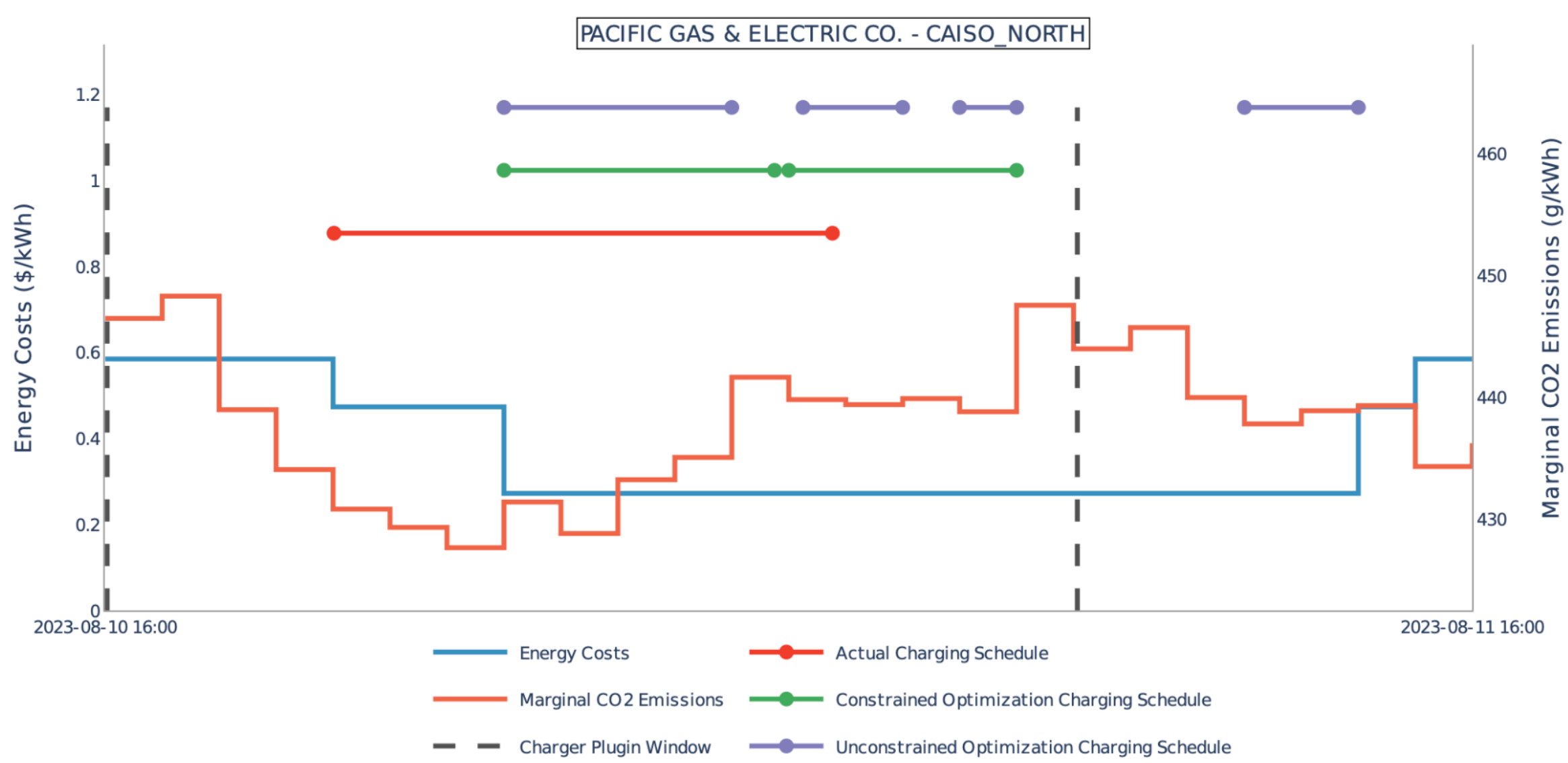

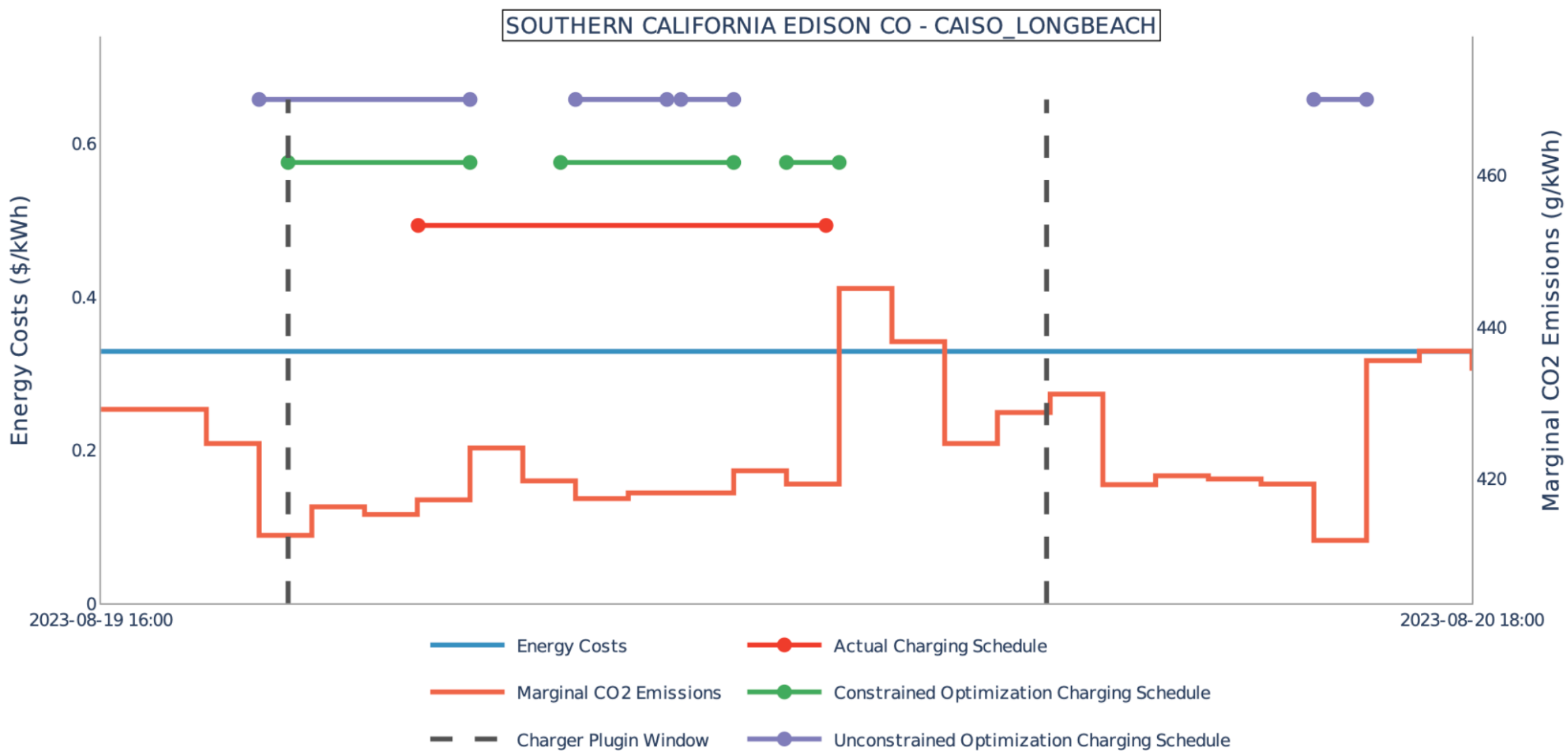

***Figure 3:*** *Illustrative Examples to demonstrate the difference between two different optimization frameworks used in the study (a) Pacific Gas & Electric Co with a TOU rate structure (b) Southern California Edison Co with a flat rate structure*

Both frameworks prioritize first minimizing cost and second minimizing carbon emissions. The optimization assumes perfect information in terms of marginal carbon emissions to assign the optimal charging times for the session. For utilities with a TOU rate structure, charging times are assigned within the lowest cost time periods, and subsequently to the lowest marginal emissions period within that same window. For utilities without a TOU rate structure, reducing the associated carbon emissions is the sole optimization, which is evident in the SCE example illustrated in Figure 3.

**Calculating Energy Costs and Carbon Emissions:**

The study applied both the constrained and unconstrained optimization frameworks to charging session data across 5,195 Rivian consumer vehicles. While the unconstrained optimization also leverages the plug-in session data, both approaches calculated the optimal charging schedules to minimize energy costs and carbon emissions. This optimal charging schedule is then used to calculate the theoretical associated energy costs and carbon emissions. These "optimal" results are then compared to the baseline energy costs and carbon emissions of the actual observed charging.

The optimization methodology assumes that customers opted into an EV-specific tariff, when available. The baseline calculations assumed the customer was on the standard residential tariff, and not the EV-specific tariff. In locations where EV-specific tariffs are available, savings calculations include both the impact of switching tariffs, and the charging schedule optimization. The percent reduction in energy costs and carbon emissions was calculated based on the difference between total baseline and optimized values, divided by the baseline. To simplify the analysis and reduce computational complexity, the study averaged WattTime's 5-minute marginal carbon emissions signals to hourly intervals.

## Results & Discussion

We observed an average 21.5% potential reduction in energy costs associated with charging per vehicle across the 58% of vehicles in locations with TOU rate structures (3,010 of 5,195). This same subset of vehicles showed a 3.9% potential decrease in carbon emissions when optimized. We found emissions reductions averaging 7.4% per vehicle for the 42% of vehicles located in areas without TOU rate structures. This relatively higher decrease in emissions can be attributed to the optimization only occurring on marginal carbon emissions. The results show that the savings potential (cost and carbon) can be realized without changing customer behavior.

Most sessions experienced reduced energy costs, however, approximately 1 in 6 sessions saw increased costs compared to the baseline across the 58% of vehicles under TOU rates. We found cost increases occurred when charging/plug-in windows overlapped with the tariff's on-peak hours. As previously noted, EV-specific utility tariffs typically offer lower rates during off-peak hours (majority of the day) while imposing higher costs during on-peak periods. While a subset of sessions resulted in higher costs post-optimization, charging costs overall were reduced by 21.5% when optimized.

***Table 2:*** *Summary of cost savings and emissions reductions*

| | **Vehicles with TOU rates** | **Vehicles without TOU rates** |
|---|---|---|
| **Number of vehicles in sample** | 3,010 | 2,185 |
| **Constrained Optimization: Cost Savings** | 21.5% | - |
| **Unconstrained Optimization: Cost Savings** | 29% | - |
| **Constrained Optimization: Carbon Emissions Reductions** | 3.9% | 7.4% |
| **Unconstrained Optimization: Carbon Emissions Reductions** | 10.1% | 15.3% |

PG&E customers had the highest potential to reduce emissions while also minimizing energy costs – 7.4% on average. In contrast, Arizona Public Service Co. and Ameren Illinois customers had the lowest potential emissions reductions, at 0.6% and 1.5% respectively, when cost minimization was the primary goal. This highlights the varying effectiveness of EV tariffs in aligning TOU periods with periods of low carbon emissions on the grid. Among customers with flat rate structures and no cost reduction potential, those served by Public Service Co. of Colorado, had the highest potential reduction in emissions at 17.4%. This can be attributed to the high carbon variability in the local grid.

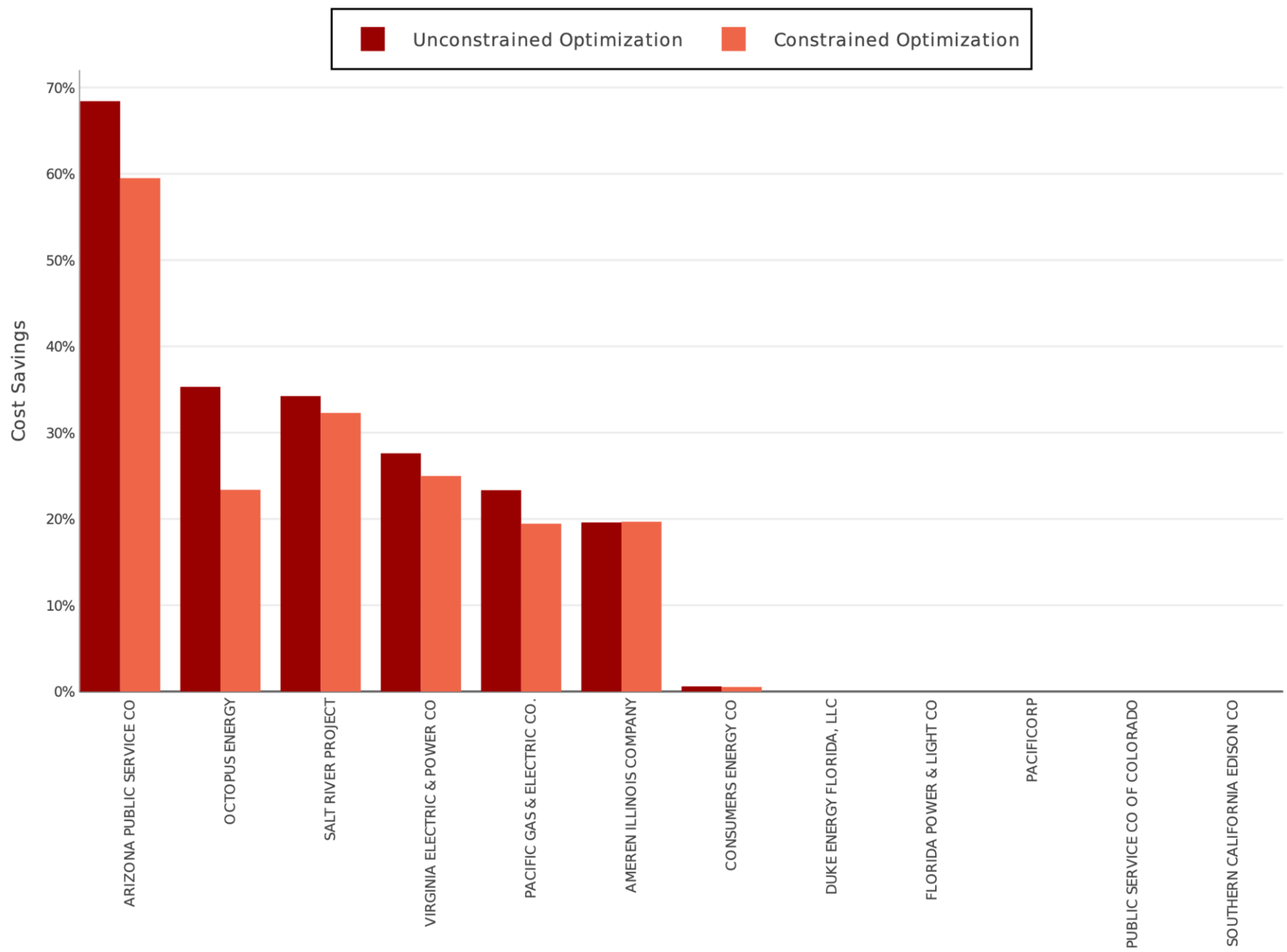

***Figure 4:*** *Cost Savings by each electric utility for two different optimization frameworks*

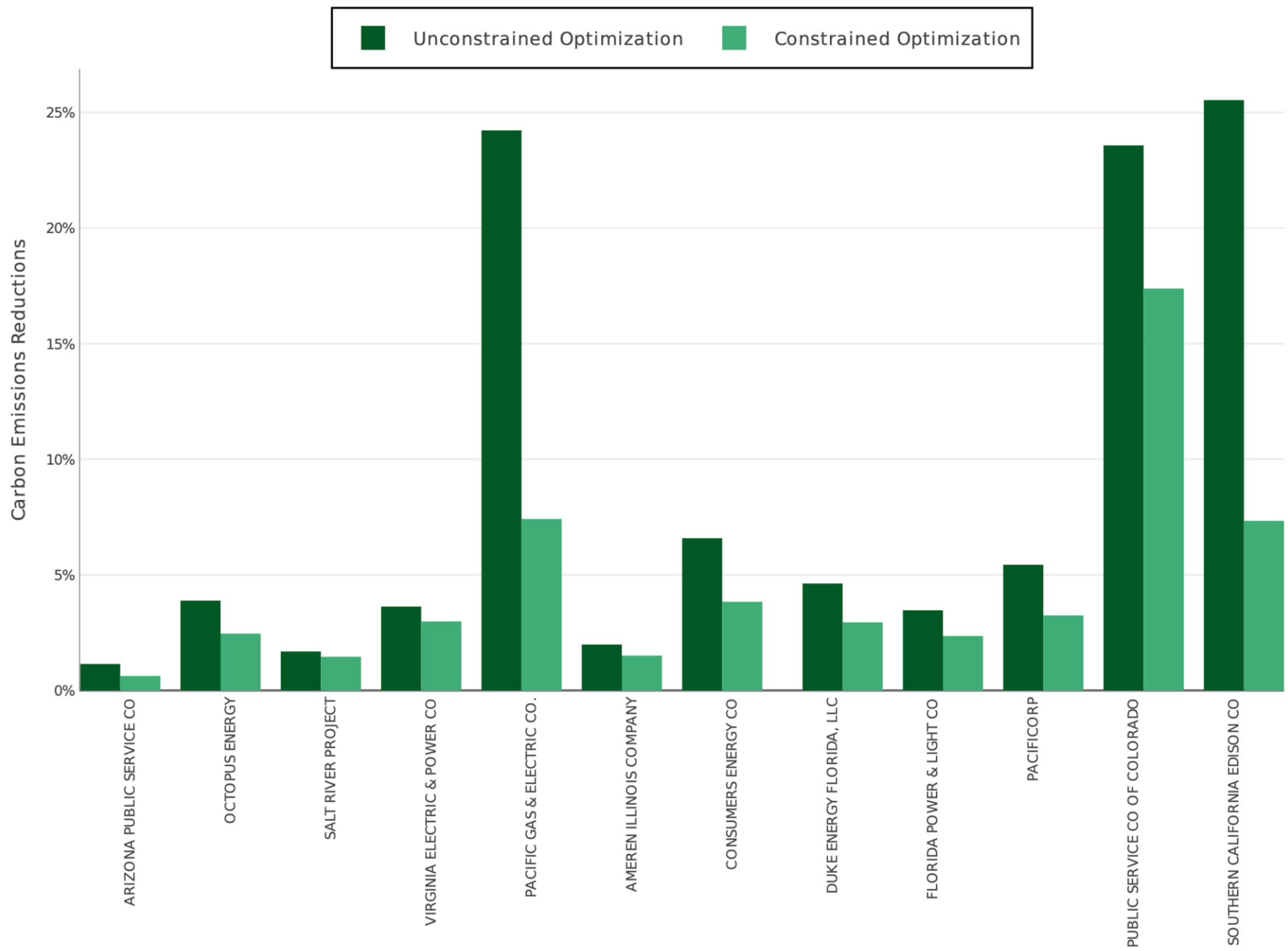

***Figure 5:*** *Carbon Emissions Reductions by each electric utility for two different optimization frameworks*

When optimizing charging schedules within the entire day for each charging session, without plugin time constraints, we observed a 29% average potential reduction in charging costs per vehicle and a 10.1% potential decrease in emissions associated with charging at home for vehicles with TOU rate structures. The remaining fleet, served by utilities offering only flat rate structures, experienced a 15.3% average potential reduction in emissions, but no cost reductions. The *unconstrained* results demonstrate an upper bound of achievable savings, if drivers modified their charging behavior to align perfectly with optimal times based on these signals, but neglected customer preferences or vehicle needs.

By contrast, the *constrained* results represent the proportion of maximum achievable savings without requiring owners to change plug-in behavior. Without changes in plug-in times, and based solely on shifting to times when charging occurs within historical actual plug-in windows, EV owners achieved significant benefits. For those vehicles with TOU rates, the constrained approach delivered 21.5% cost savings, compared to a maximum potential of 29% in the unconstrained case. Additionally, a vehicle achieved a 3.9% reduction in emission in the constrained case versus a 10.1% maximum observed in the unconstrained case. For flat rate customers, the constrained approach reduced emissions by 7.4%, compared to a 15.3% maximum potential emissions reduction in the unconstrained case. Our findings demonstrate that EV owners realized most of the cost savings and about half of the emissions reduction potential, all within their existing routines. Even greater emissions and cost savings are achievable by adjusting vehicle charging windows when possible.

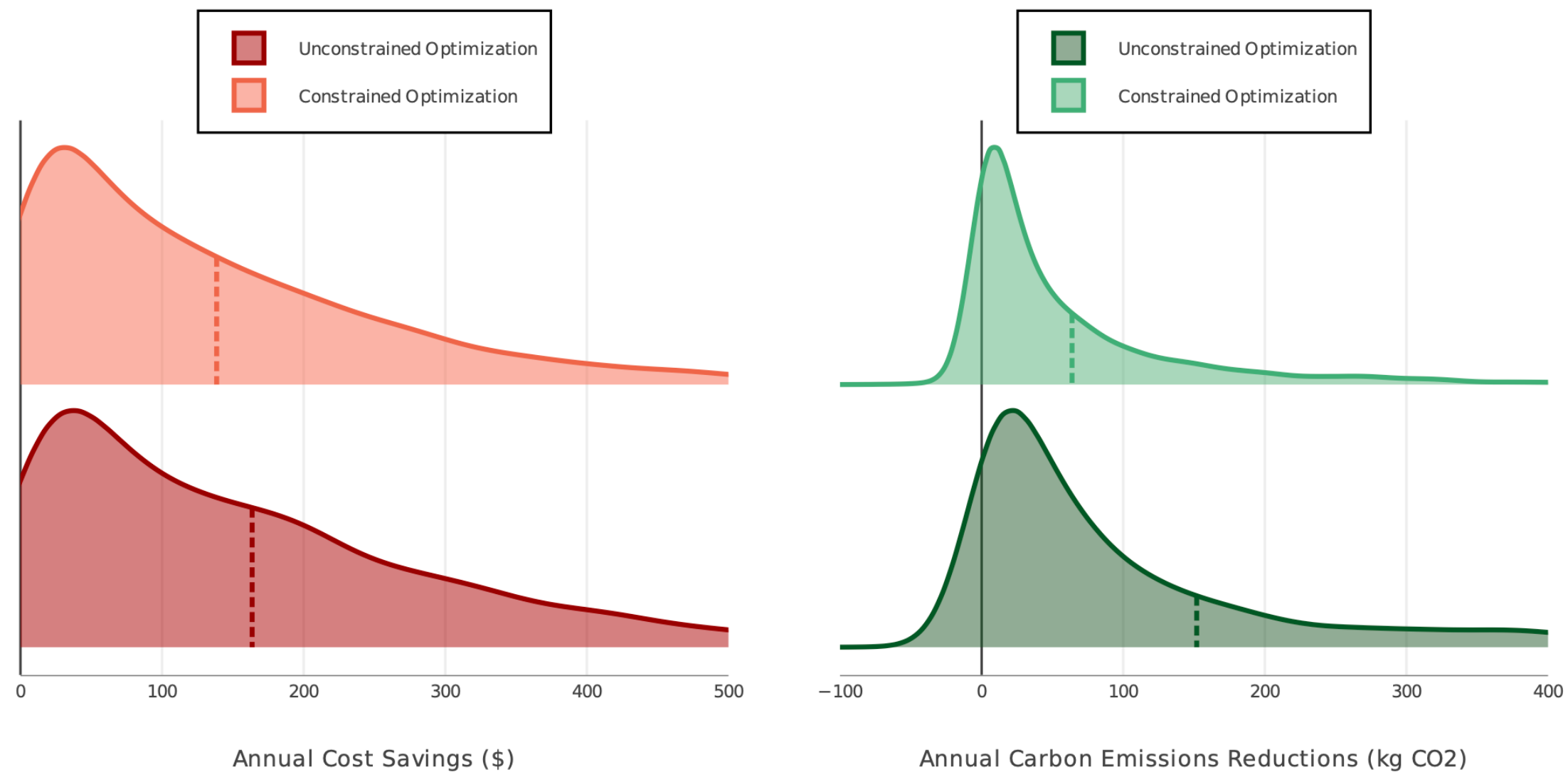

***Figure 6:*** *a) Annual Cost Savings (mean = $164 for unconstrained, $139 for constrained; median = $124 for unconstrained, $100 for constrained) and b) Carbon Emissions Reductions (mean = 152 kg for unconstrained, 64 kg for constrained; median = 63 kg for unconstrained, 28 kg for constrained) per vehicle for two different optimization frameworks (*only vehicles with TOU rate are used in cost savings box plot)*

Figure 6 illustrates the distribution of annual cost savings and carbon emissions reductions per vehicle/customer, comparing the two optimization frameworks employed in this study. Without altering their lifestyle, EV owners can save up to $900 annually, with an average savings of approximately $140, and a significant reduction in their carbon footprint by utilizing smart charging.

## Conclusions

This study confirms with historical charging telematics data, provided by Rivian Automotive, that smart charging has significant potential to reduce both electricity costs and carbon emissions from residential EV charging. By aligning charging with time periods when the grid is underutilized, drivers can achieve substantial savings without altering their driving habits. The analysis reveals that both the constrained and unconstrained charging scenarios yield cost and carbon reductions, highlighting the effectiveness of smart charging in various use cases. Notably, when on a TOU rate and without changing plug-in behavior, drivers can still achieve meaningful savings and contribute to a more efficient and cleaner grid. Although not the primary focus, this study also sheds light on the complex relationship between tariff structures and grid carbon intensity. Different resource mixes and demand profiles across regions mean that a one-size-fits-all approach to incentivizing charging behaviors to reduce cost and emissions may not always be effective achieving both outcomes, highlighting the need for careful consideration of regional energy characteristics when designing utility tariffs and promoting smart charging.

It is important to acknowledge that the study relies solely on data from Rivian vehicles charging throughout 2023. Given the demographic concentration of Rivian customers, these data are unlikely to fully capture the diversity of EV ownership, representative charging behaviors across all EV owners, and customer cost sensitivity across the entire sector. Additionally, the geographic scope of the study, while encompassing various regions, may not fully represent the nuances of electricity pricing and grid conditions in all areas - Rivian is a new EV company and subsequently has a high concentration of vehicles located in California. Furthermore, the moving variables of consistent charging habits and shifts in energy market structures, and vehicle-grid integration could impact incremental effectiveness. Despite these limitations, the findings highlight the importance of future research and development to include a wider range of EV models, geographic locations, and charging habits. Further investigation into the impact of dynamic pricing models and the changing energy mix will be crucial in maximizing grid efficiency. The study does not test methods for helping customers shift charging times automatically to align with cost or carbon optimization. We see this as another promising area of research to pursue as a strategy to help customers realize the theoretical savings identified in this study.

To maximize the benefits of smart charging, a multi-pronged approach is crucial. Educating EV owners on its benefits can empower them to save money and drive progress towards a net-zero future. Rivian and other EV manufacturers can play a key role by integrating smart charging platforms that incorporate grid signals, increase adoption and utilization of TOU rates, and simplify the user experience through automated and dynamic scheduling. Furthermore, utilities and policymakers can incentivize adoption by offering more granular and transparent rate structures and further empowering consumers with tools to manage their energy usage. Designing rate structures and software tools that encourage beneficial charging behaviors can empower EV owners to save money while actively contributing to a cleaner, more sustainable and cost-efficient energy future.

## Data Availability

The raw charging session data used in this study is not publicly available to protect consumer privacy. This data is collected from Rivian customers who opted-into data sharing for analytics purposes, and access is restricted to internal use to maintain user confidentiality. However, the aggregated distribution of plugin and charging sessions by time-of-day and duration bins is available at https://doi.org/10.5281/zenodo.17353155. This distribution data underpins the behavioral patterns visualized in Figures 1 and 2 of the main paper, showing typical vehicle plugin times to chargers at home and their session durations. Each cell in the dataset represents the count of sessions. For "Plugged In" rows, the values are cumulative, representing sessions lasting at least that duration. The "Charging" rows show sessions actively charging within each duration window, grouped by the plugin start time of day. The distribution uses 30-minute granularity for both time-of-day (48 buckets) and duration bins (up to 48 hours), preserving usage patterns while protecting the privacy of Rivian customers. Researchers can use this distribution data to replicate our findings on EV owners' home charging behavior, analyze patterns, and generate synthetic datasets for further research in smart charging optimization and grid impact studies.

The utility tariff rate information was accessed through Arcadia's Signal platform, which is available for commercial use, but the same data is also publicly accessible on individual utility websites. The marginal carbon emissions data was provided by WattTime and is available for commercial use.

### Code Availability

The code used in this study can be made available upon reasonable request. However, access may be limited to protect consumer privacy and confidentiality of the underlying data, as it contains processing methods specifically designed for sensitive customer information.

## Acknowledgements

The authors would like to thank WattTime for their valuable input and access to data resources. WattTime is an environmental tech nonprofit that empowers all people, companies, policymakers, and countries to slash emissions and choose cleaner energy. The authors declare no conflicts of interest and are employees of Rivian Automotive with no personal or financial relationships that could bias the research findings.

## Author Contributions

Yash Gupta contributed to the study design, conducted the analysis, created the visualizations, and wrote the original manuscript. William Vreeland, Andrew Peterman, and Coley Girouard contributed to the study design and participated in the review and editing of the manuscript. Brian Wang contributed to the study design and reviewed the manuscript. All authors approved the final version and agreed to be accountable for the integrity of the work.

## Competing Interests

The authors declare no competing interests.

## Supplementary Information

**Rivian vehicles and Charging/Plugin Sessions Data:**

This study analyzed charging session data from 5,195 Rivian vehicles, encompassing over 432,900 sessions throughout 2023 (January 1, 2023 – December 31, 2023). This data represents charging sessions by Rivian customers who own a dedicated home charging station supplied by Rivian known as a Wall Charger and who were opted-into sharing certain vehicle data for product research and development. Further, we selected the data for customers served by some of the largest utilities by coverage region in the United States. We also filtered to only keep utilities that served at least 100 Rivian customers, ensuring that the sample size is large enough to yield reasonable and statistically significant conclusions for each utility. This led to a final sample set of Rivian customers served by 12 major utilities spanning across 13 different states in the United States as depicted in Supplementary Figure 1 & 2.

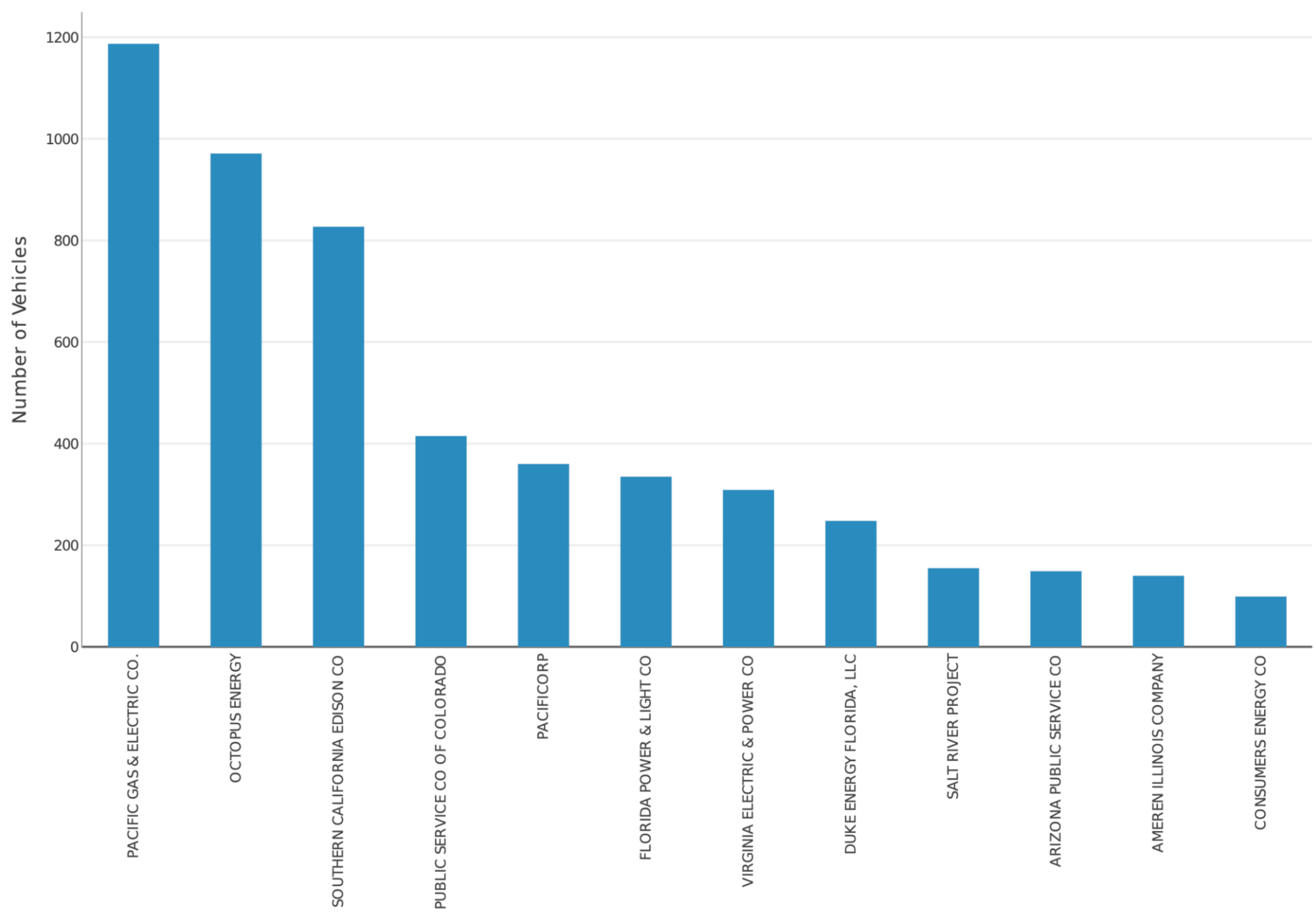

***Supplementary Figure 1:*** *Number of Rivian vehicles by each Electric Utility*

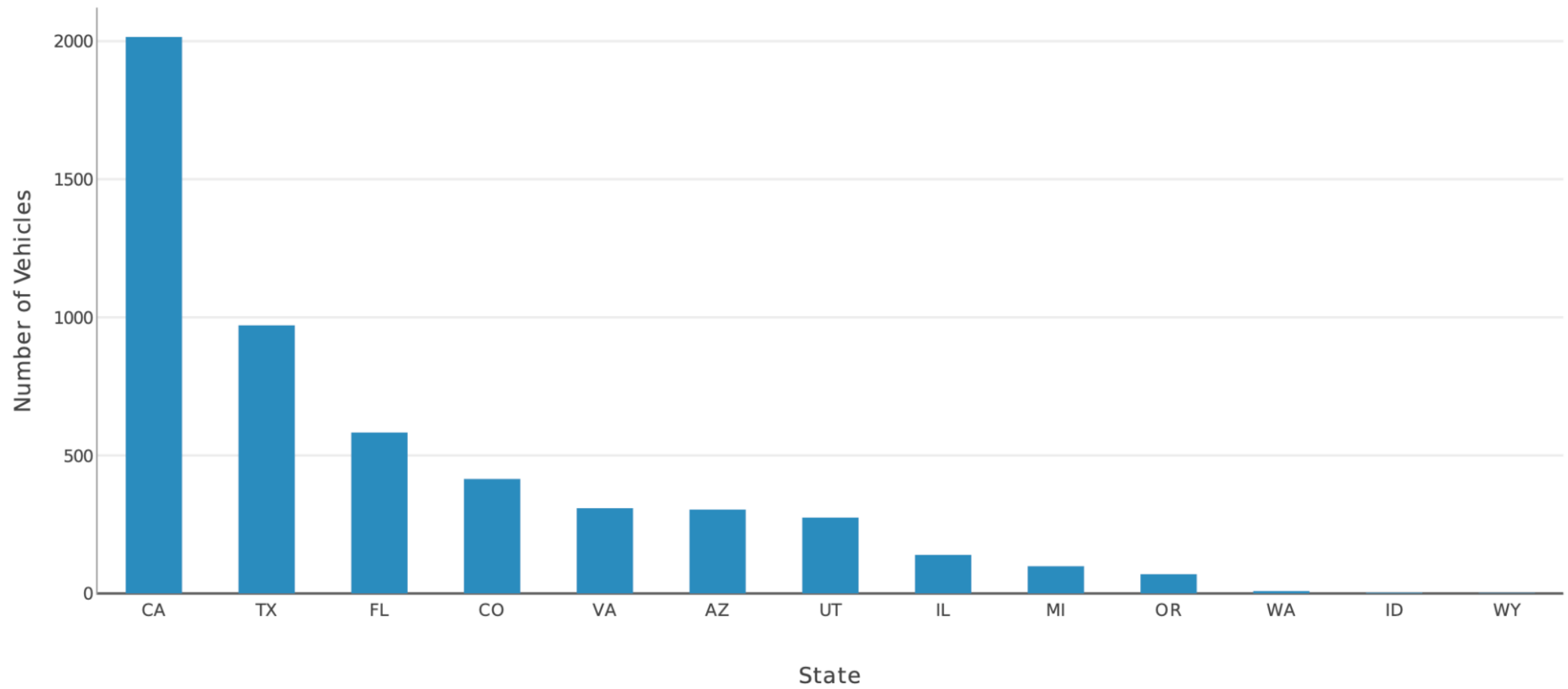

***Supplementary Figure 2:*** *Number of Rivian vehicles by each state*

Supplementary Figure 3 shows the distribution of charging sessions per vehicle during the study period, revealing that the median Rivian owner charges 126 times annually (mean 138) at home. Supplementary Figure 4 presents the distribution of home charging energy for Rivian owners. It is important to note that out of 5,195 vehicles included in the study, roughly 2,000 were delivered before the beginning of 2023 and have an entire year of data. The remainder of the vehicles were delivered partway through the study period with an evenly distributed delivery date across the year, resulting in fewer charging sessions and lower energy usage on a per-vehicle basis compared to the rest of the fleet for vehicles that were delivered later in the year.

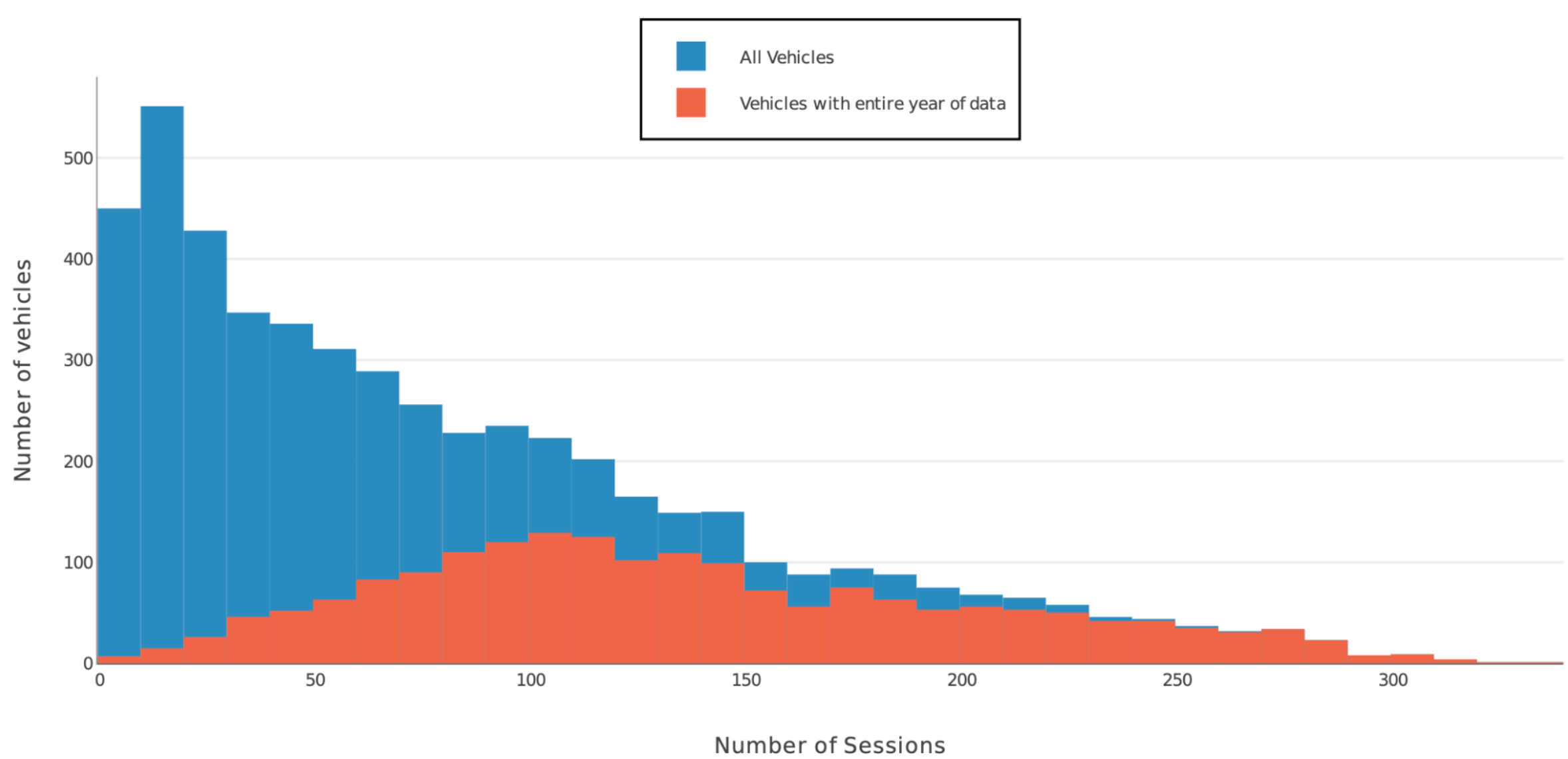

***Supplementary Figure 3:*** *Distribution of number of sessions per vehicle in 2023 (mean = 138; median = 126 for vehicles with entire year of data)*

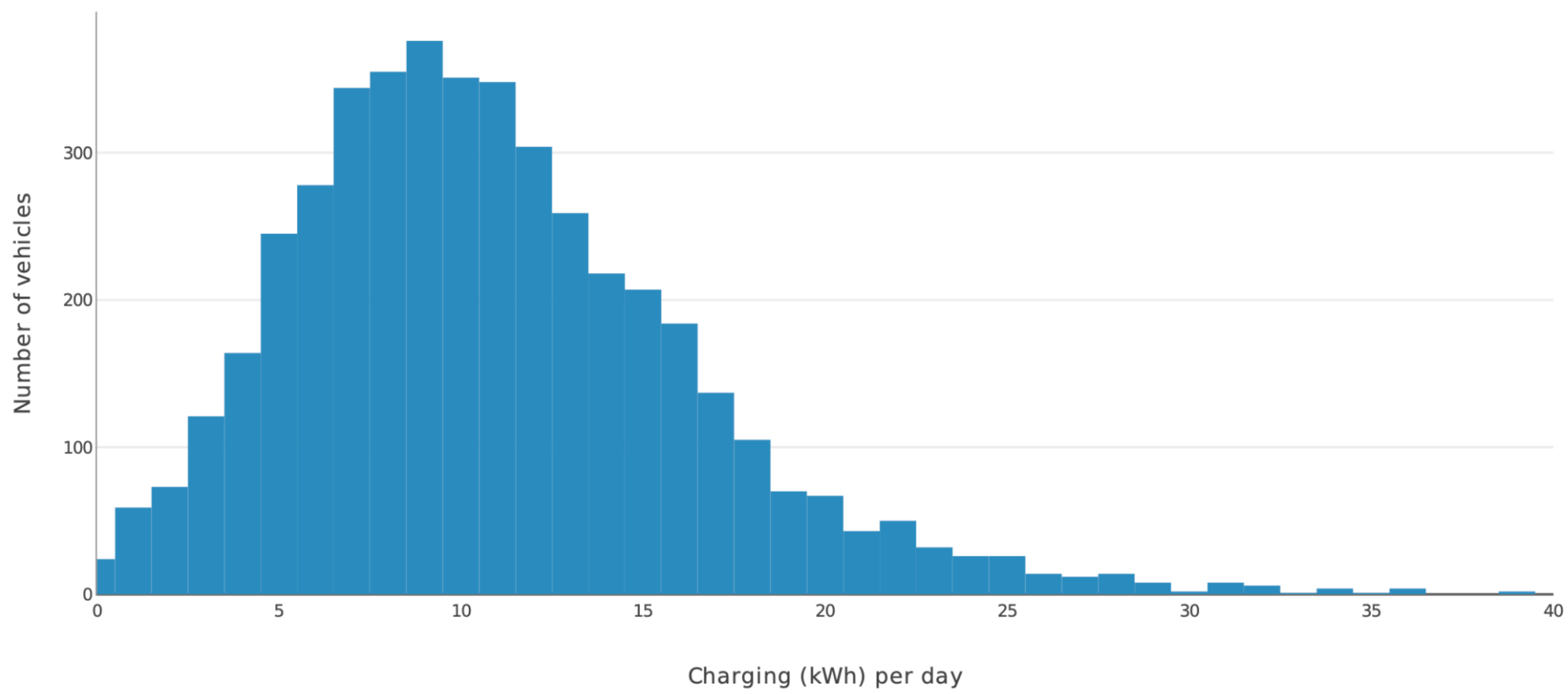

***Supplementary Figure 4:*** *Distribution of charging energy per day per vehicle (mean = 10.9 kWh; median = 10.1 kWh)*

Supplementary Figure 4 shows the distribution of the amount of charging done by each vehicle included in the study on a day-to-day basis. Note that the daily charging energy for a vehicle is independent of when that vehicle was delivered.

**Utility Tariffs:**

This research utilizes residential utility tariffs data from Arcadia. Supplementary Table 1 presents the residential tariffs, as well as EV-specific tariffs where available, for the utilities included in this study. These rate structures not only vary by utility but also differ by state, particularly for utilities that operate across multiple states. Additionally, electric utilities often offer residential customers multiple tariff options with different rate structures. Many tariffs follow flat rate structures while others have varying costs based on TOU, consumption tiers, seasons, or a combination thereof. The majority of utility customers are registered on the base residential tariff offered by default. We found this base rate (either default or most commonly selected tariff) offered by each utility in the states covered by them (as listed in Supplementary Table 1) and used that to calculate the baseline costs.

***Supplementary Table 1****: Electric Utilities along with their residential tariffs*

| Utility | State | Residential Base Tariff | Base TOU Rate Structure? | EV Tariff | EV TOU Rate Structure? |
|---|---|---|---|---|---|
| PACIFIC GAS & ELECTRIC CO. | CA | E-1 Residential | NO | EV-2A-TOU Residential - Time of Use - Plug-In Electric Vehicle 2 | YES |
| SOUTHERN CALIFORNIA EDISON CO | CA | Domestic Code-D | NO | | |
| PACIFICORP | OR | Residential Service Code-4 | NO | Residential - Electric Vehicle Code-5 | NO |
| PACIFICORP | WA | Residential Code-16 | NO | | |
| PACIFICORP | CA | Residential Code-D | NO | | |
| PACIFICORP | UT | Residential Code-1 | NO | | |
| PACIFICORP | ID | Residential Code-1 | NO | | |
| PACIFICORP | WY | Residential Code-2 | NO | | |
| PUBLIC SERVICE CO OF COLORADO | CO | Residential Code-R | NO | | |
| FLORIDA POWER & LIGHT CO | FL | RS-1 Residential | NO | | |
| DUKE ENERGY FLORIDA, LLC | FL | RS-1 Residential | NO | | |
| AMEREN ILLINOIS COMPANY | IL | DS-1 Residential | NO | DS-1-EVCP Residential - Electric Vehicle Charging Program | YES |
| SALT RIVER PROJECT | AZ | E-23 Residential | NO | E-29 Residential - Experimental Time-Of-Use, Super Off-Peak - Electric Vehicle | YES |
| CONSUMERS ENERGY CO | MI | RSP Residential-Summer On-Peak | YES | | |
| VIRGINIA ELECTRIC & POWER CO | VA | Residential(VA) Code-1 | NO | 1EV Residential - Electric Vehicle | YES |
| ARIZONA PUBLIC SERVICE CO | AZ | R-1 Residential - Fixed Energy Charge Plan, Medium Tier | NO | R-EV Residential - Electric Vehicle Time-Of-Use | YES |
| Octopus Energy | TX | Residential Octo Base | NO | Residential Octo Riv EV | YES |

In the residential customer class, an increasingly common option is an EV tariff. Residential customers with a registered electric vehicle are eligible for this special EV tariff if offered by their utility. Based on the utility tariff database provided by Arcadia, as of Dec 2024, there are 71 utilities that offer these EV-specific tariffs across 28 states in the US. This number has increased compared to only 54 residential EV tariffs in April 2023 [1] and we can expect more utilities to launch their own EV tariffs in the near future. EV tariffs typically feature TOU rates, which consist of pre-defined peak and off-peak time periods with a tiered structure for each, usually offering lower off-peak prices to encourage EV charging during those hours. As with other TOU tariffs, EV tariffs are generally structured to incentivize customers to charge at times that reduce grid and customer costs. While TOU once seemed like a major departure from traditional fixed/flat rate designs, advancements in technology have made it easier for customers to adopt, providing real-time insights and tools to respond to market signals. Early adopters have shown that EV users are ready to embrace this shift. However, there is a downside for customers charging during on-peak hours on an EV specific tariff.

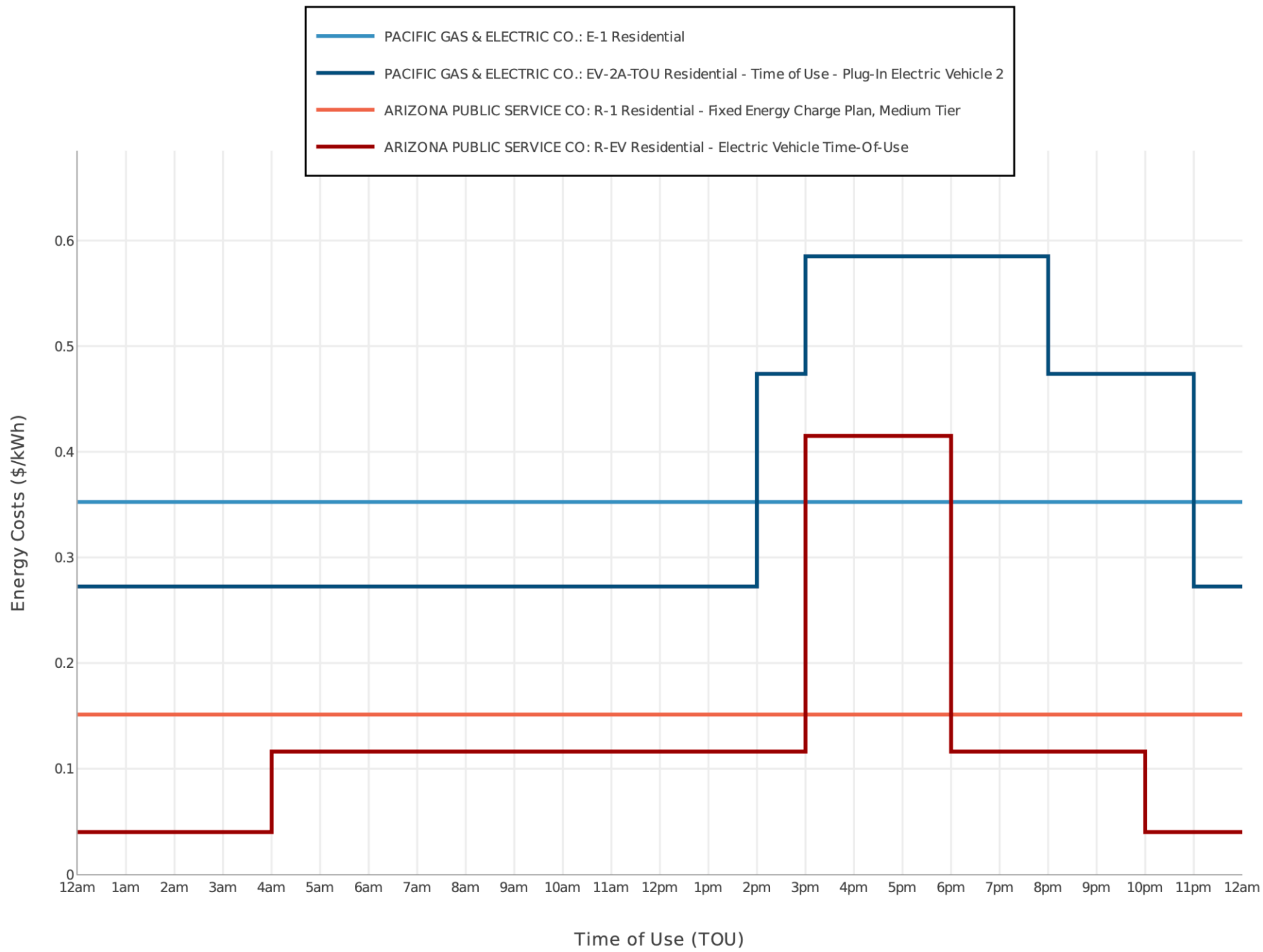

***Supplementary Figure 5:*** *Sample residential tariffs and EV tariffs for PG&E and APS for a sample month (July 2023)*

Supplementary Figure 5, which shows the residential and EV-specific tariffs offered by PG&E (in California) and Arizona Public Service Co. (in Arizona), highlights the potential cost savings for customers who switch to an EV tariff and charge during off-peak hours. However, as shown in Supplementary Table 1, not all electric utilities currently offer TOU or EV-specific tariffs. For customers on tariffs with flat rate structures, where no cost savings are available, there is still an opportunity to reduce the carbon emissions associated with their residential charging. In these cases, the study focuses on optimizing marginal carbon emissions.

**Marginal Carbon Emissions Data:**

The importance of using marginal emissions factors rather than average emissions factors for smart charging applications has been demonstrated by Huber, J. et al. [2], showing that average factors can lead to increased emissions during charging periods. This analysis used marginal operating emissions rates (MOERs) provided by WattTime to determine the emissions-optimized charging schedules and their potential emission savings. WattTime calculates marginal emissions rates using a regression model that finds the relationship between total emissions and load with causal confounding variables controlled. The regression model is trained on actual, not assumed, grid data to produce a change in emissions given a change in supply or demand. Causal confounding variables are controlled so that changes in emissions measured by the model can be consequentially attributable to the change in supply or demand instead of due to unrelated phenomenon (such as changes in temperature) [3].

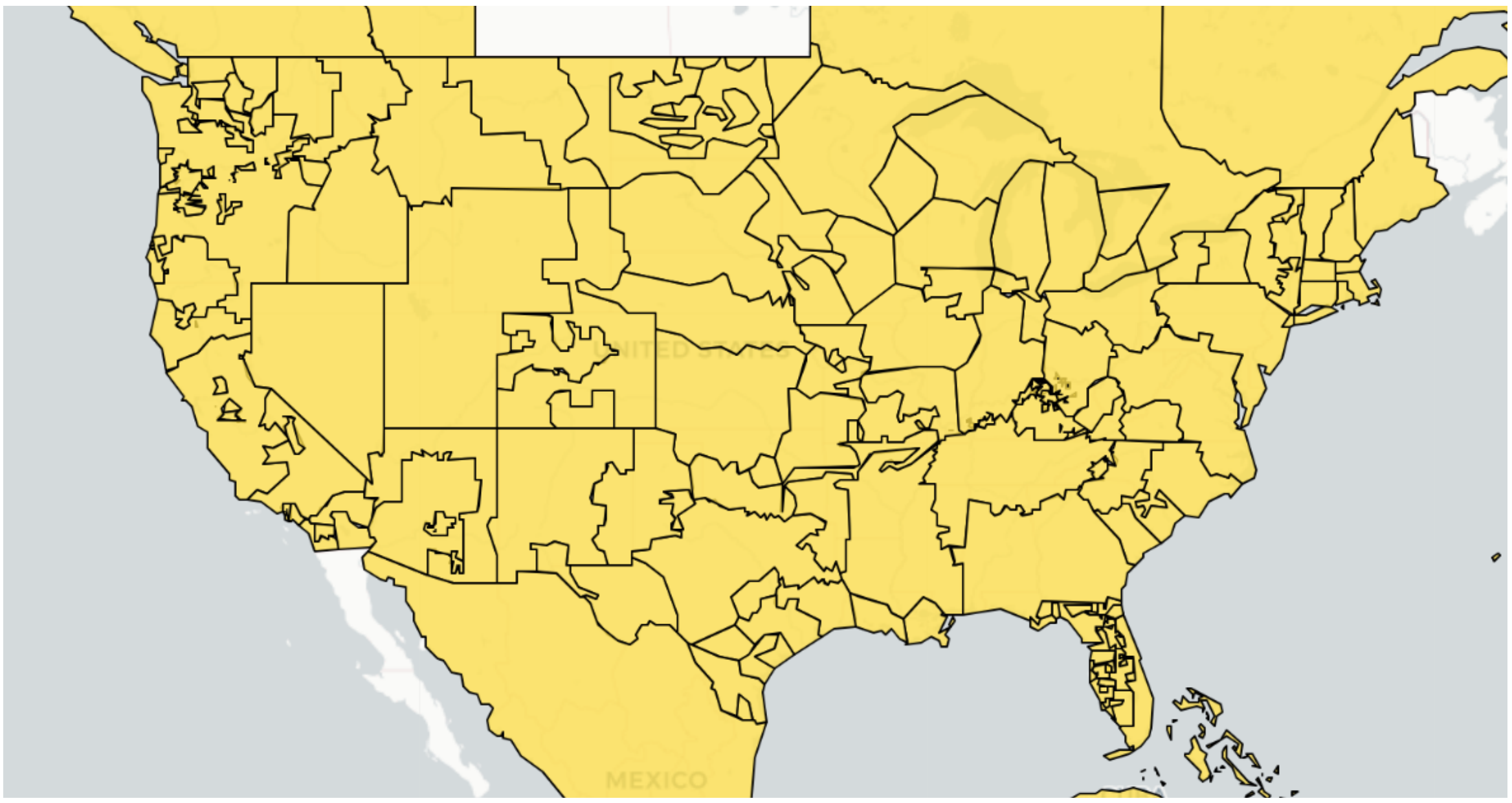

***Supplementary Figure 6:*** *US grid boundaries where WattTime provides both historical and forecasted marginal carbon emissions signal*

WattTime publishes historical, real-time, and forecasted 5-min marginal carbon emissions, for over 210 countries and territories around the world and makes this available via an API that can be integrated into the control systems of electricity consuming devices to shift load to low emissions periods. Supplementary Figure 6 shows the grid regions within the US reflecting balancing authorities and ISO subregions.

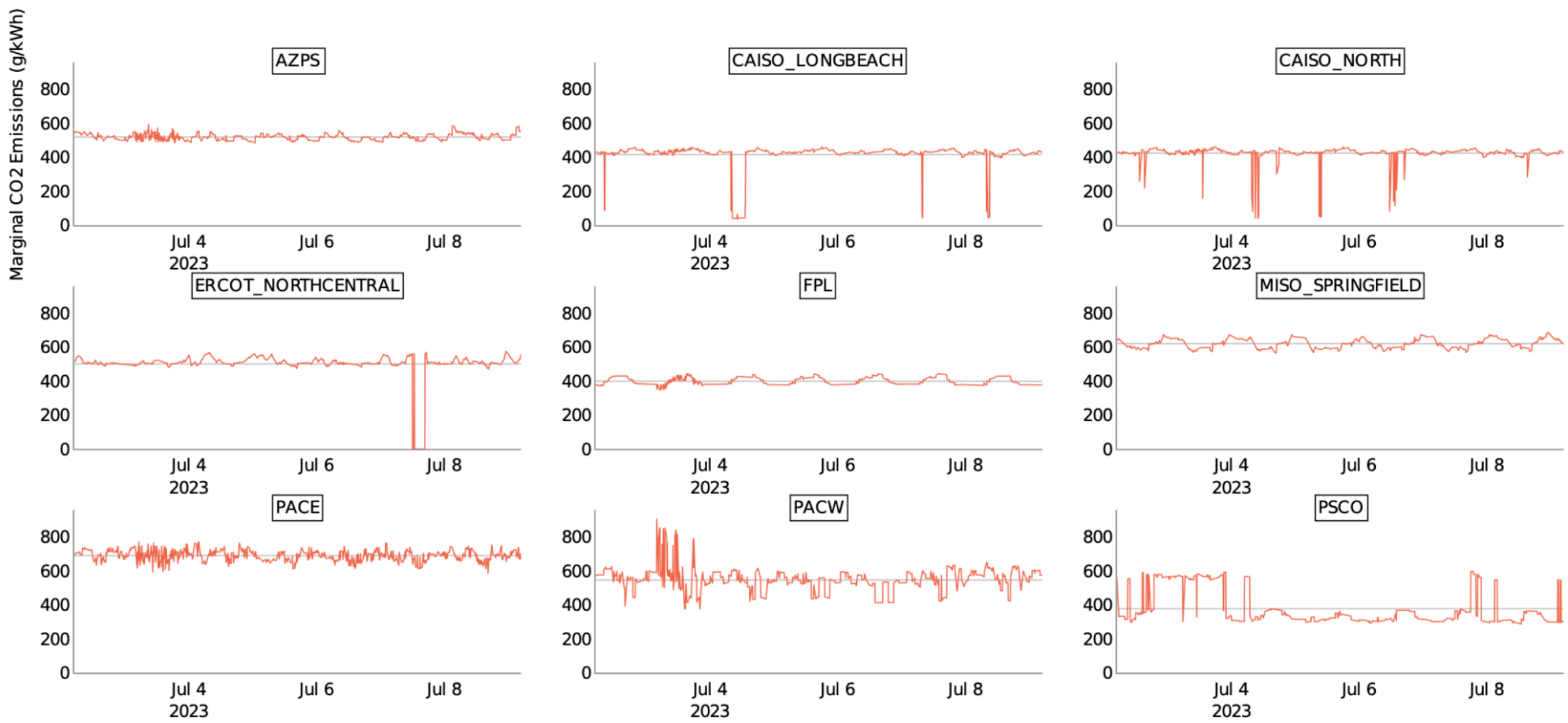

***Supplementary Figure 7:*** *Marginal Carbon Emissions of a sample week in 2023 for the grid regions included in the study*

The marginal emissions rates in a grid region are driven by the types of generators responding to changes in load. Grid regions with renewable energy curtailment - a situation in which excess renewable energy is discarded - tend to have increased variability in marginal carbon emissions. For example, in the California Independent System Operator (CAISO) grid region, the marginal carbon emissions can swing from ~0 g CO2/kWh when solar energy is being curtailed, to ~400 g CO2/kWh when a natural gas plant is supplying additional electricity to the grid. Conversely, grid regions like Arizona Public Service Co, don't have a lot of variability in marginal carbon emissions rates because the primary marginal power plants in those regions are coal or natural gas.

**Utility Tariff Rates vs Marginal Carbon Emissions:**

With optimizing costs as a primary objective, the carbon emissions associated with the charging session might actually go up for vehicles served by utilities with negative correlation between energy rates and marginal carbon emissions. For example, in regions heavily reliant on fossil power plants, like some parts of the Midwest, shifting charging to off-peak hours (often at night) might lower electricity costs due to lower demand, but could actually *increase* carbon emissions. This is because baseload power plants often run continuously overnight, and shifting demand to these hours increases their utilization. However, in regions like California, where solar power is abundant during the day, shifting charging to these daytime hours can both lower costs and reduce reliance on fossil fuel peaker plants that are often used to meet evening demand. This alignment of cost and carbon reduction occurs because the lowest cost electricity (solar) is also the cleanest. Therefore, while cost optimization is important, it must be carefully considered alongside regional generation mixes and demand patterns to ensure that cost savings translate into carbon reductions.

***Supplementary Table 2:*** *Correlation between various utility rates and corresponding grid's marginal carbon emissions by time of day in July 2023*

| Electric Utility | Grid Region | Correlation between energy costs and marginal carbon emissions |
|---|---|---|
| Pacific Gas & Electric Co | BANC | 87% |
| Virginia Electric & Power Co | PJM – DC | 83% |
| Pacific Gas & Electric Co | CAISO – North | 79% |
| Salt River Project | SRP | 54% |
| Arizona Public Service Co | AZPS | -1% |
| Consumers Energy Co | MISO – Grand Rapids | -52% |
| Ameren Illinois Company | MISO – Springfield | -86% |

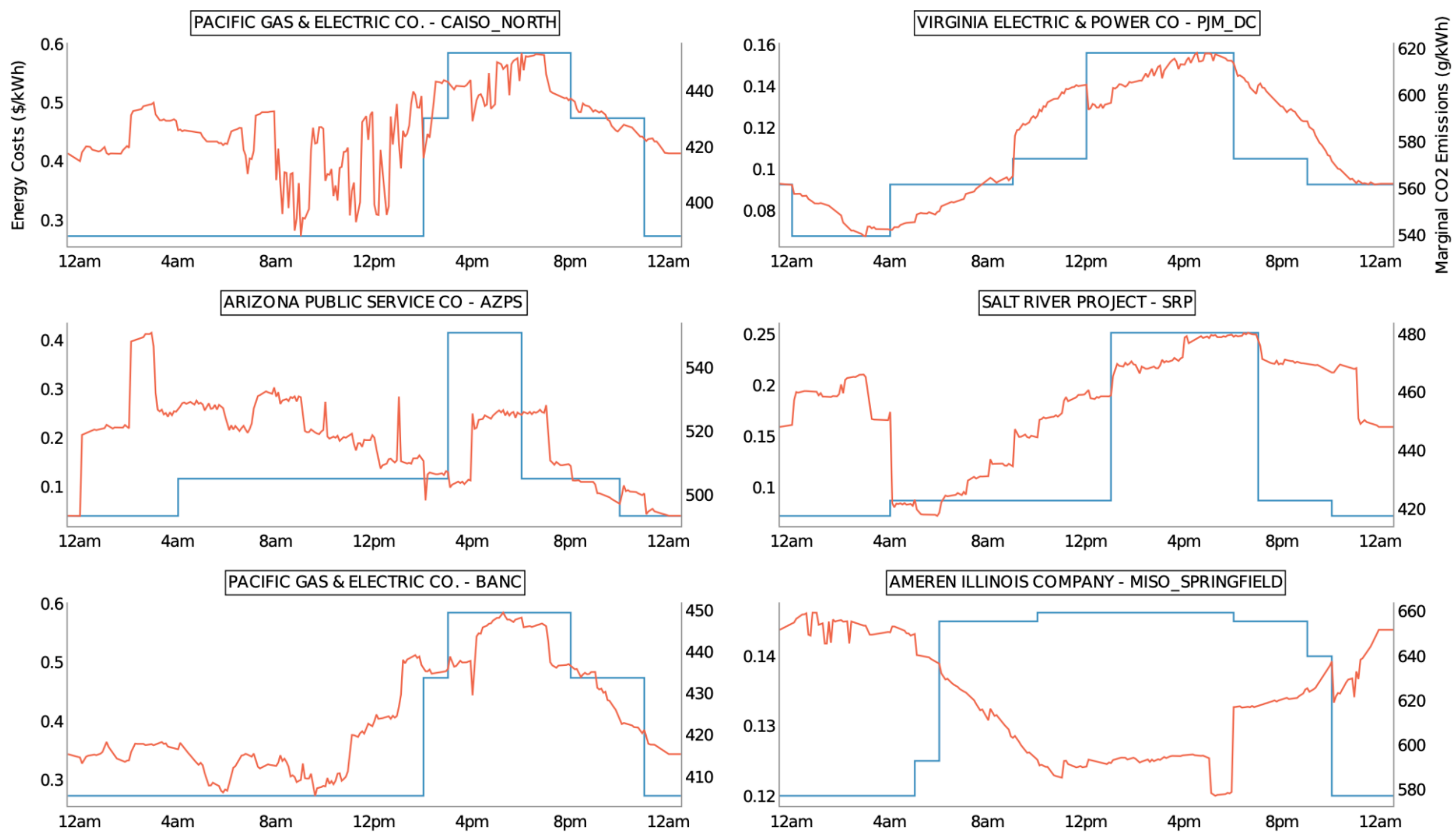

***Supplementary Figure 8:*** *A comparison of utility rates against marginal carbon emissions for various electric utilities and grid regions in July 2023*